\documentstyle{article}
\begin{document}

\title{ 
Is the Two-Dimensional One-Component Plasma Exactly Solvable?
}

\author{
L. {\v S}amaj\footnote{
Institute of Physics, Slovak Academy of Sciences,
D\'ubravsk\'a cesta 9, 845 11 Bratislava, Slovakia;
e-mail: fyzimaes@savba.sk}
}

\maketitle

\begin{abstract}
The model under consideration is the two-dimensional (2D) one-com\-po\-nent
plasma of pointlike charged particles in a uniform neutralizing background,
interacting through the logarithmic Coulomb interaction.
Classical equilibrium statistical mechanics is studied 
by non-traditional means.
The question of the potential integrability (exact solvability)
of the plasma is investigated, first at arbitrary coupling constant 
$\Gamma$ via an equivalent 2D Euclidean-field theory, and then at 
the specific values of $\Gamma=2*$integer via an equivalent 1D 
fermionic model.
The answer to the question in the title is that there is strong evidence 
for the model being not exactly solvable at arbitrary $\Gamma$ 
but becoming exactly solvable at $\Gamma=2*$integer.
As a by-product of the developed formalism, the gauge invariance
of the plasma is proven at the free-fermion point $\Gamma=2$;
the related mathematical peculiarity is the exact inversion of a class 
of infinite-dimensional matrices.
\end{abstract}

\medskip


\newpage

\renewcommand{\theequation}{1.\arabic{equation}}
\setcounter{equation}{0}

\section{INTRODUCTION}
In this paper, we consider a classical (i.e. non-quantum) model
which belongs to the general class of two-dimensional (2D) Coulomb
systems of charged particles.
According to the laws of 2D electrostatics, the particles can
be thought of as infinitely long charged lines which are
perpendicular to the confining surface. 
Thus, the electrostatic potential $v$ at a point ${\bf r}$, 
induced by a unit charge at the origin, is given by
the 2D Poisson equation
\begin{equation} \label{1.1}
\Delta v({\bf r}) = - 2 \pi \delta({\bf r})
\end{equation}
In a plane, the solution of this equation, subject to the boundary
condition $\nabla v({\bf r}) \to 0$ as $\vert {\bf r} \vert \to \infty$, 
reads
\begin{equation} \label{1.2}
v({\bf r}) = - \ln \left( \frac{\vert {\bf r}\vert}{r_0} \right)
\end{equation}
where the free length constant $r_0$, which fixes the zero point
of the potential, will be set for simplicity to unity.
In the Fourier space, the Coulomb potential (\ref{1.2}) exhibits
the characteristic small-$k$ behavior 
${\hat v}({\bf k}) = 1/\vert {\bf k}\vert^2$.
This maintains many generic properties (like screening and 
the related sum rules \cite{Martin}) of ``real'' 3D Coulomb
fluids with the interaction potential 
$v({\bf r}) = 1/\vert {\bf r}\vert$, ${\bf r}\in R^3$.
The pair interaction energy of particles with charges $q$ and $q'$,
localized at the respective positions ${\bf r}$ and ${\bf r}'$, is
\begin{equation} \label{1.3}
v({\bf r},q;{\bf r}',q') = q q' v(\vert {\bf r}-{\bf r}'\vert)
\end{equation}

A given continuous Coulomb system is classified via the number
$M$ of different mobile (pointlike) species $\alpha = 1, 2,\ldots, M$,  
with the corresponding charges $q_{\alpha}$ and particle densities 
$n_{\alpha}$, embedded in a fixed uniform background of
charge density $\rho_b$.
The most studied versions are the one-component plasma (OCP),
or jellium, and the symmetric two-component plasma
(TCP), sometimes called the Coulomb gas.
In the OCP there is only one mobile species, $M=1$, with
$q_1=q$ and $n_1=n$, and neutralizing background of
charge density $\rho_b$.
It is useful to introduce the background ``number density'' 
$n_b$ such that $\rho_b=-qn_b$; the neutrality condition is then
equivalent to $n=n_b$.
The symmetric TCP corresponds to $M=2$, namely $q_1=q$ and $n_1=n/2$,
$q_2=-q$ and $n_2=n/2$ ($n$ stands for the total particles density), 
with no background charge $\rho_b=0$.
Due to the logarithmic nature of the interaction,
the equilibrium statistical mechanics of the underlying 2D Coulomb
systems at the inverse temperature $\beta=1/(k_B T)$ depends 
exclusively on the dimensionless coupling constant $\Gamma = \beta q^2$;
the particle density $n$ only scales appropriately the distance.
Both OCP and TCP are solvable in the high-temperature Debye-H\"uckel
limit $\Gamma\to 0$ (in the bulk \cite{Blum} as well as for
finite systems \cite{Jancovici1,Torres}) and at the free-fermion point
$\Gamma=2$ (see reviews \cite{Jancovici2,Forrester1}). 
Through a simple scaling argument, the density derivatives
of the Helmholtz free energy can all be calculated exactly
at arbitrary $\Gamma$.
For instance, the exact equation of state for the pressure $P$,
$\beta P = n ( 1 - \Gamma/4 )$, has been known for 
a long time \cite{Salzberg}.
On the other hand, the temperature derivatives of the free energy,
such as the internal energy or the specific heat, are highly
nontrivial quantities which were obtained only in the stability region 
$\Gamma<2$ of the 2D TCP by exploring the equivalent sine-Gordon model
(for a short review, see Ref. \cite{Samaj1}).

The 2D OCP, which will be of interest in this paper, 
is formally related to the eigenvalue distribution
of certain complex random matrices \cite{Ginibre} and to the
normalization problem of the Laughlin's wave functions in the
fractional quantum Hall effect \cite{Prange,Francesco}.
There are indications from numerical simulations that, 
around $\Gamma \sim 142$, the fluid system undergoes a phase transition 
to a 2D Wigner crystal \cite{Choquard1}.
The existence of this transition has been put in doubt
in a more recent paper \cite{Moore}.
As was already mentioned, by mapping onto free fermions the model is 
exactly solvable at the coupling $\Gamma=2$, in the bulk \cite{Jancovici3}
as well as in some inhomogeneous situations \cite{Jancovici2,Forrester1}.
The other exact information comes from the sum rules for truncated
particle correlations valid at arbitrary $\Gamma$ of the fluid regime.
The usual zeroth- and second-moment conditions \cite{Martin},
having analogue in any dimension, are supplemented by the
fourth-moment (compressibility) sum rule \cite{Vieillefosse},
available explicitly due to the knowledge of the exact equation
of state, and the sixth-moment condition \cite{Kalinay1},
related to universal finite-size properties of the Coulomb system
\cite{Forrester2}-\cite{Jancovici5}.
At couplings $\Gamma=2\gamma$ ($\gamma$ a positive integer), the partition
function of the 2D OCP confined to a domain can be calculated exactly 
up to a relatively large finite number of particles $N$.
For $\gamma$ being an odd integer, the methods based on the expansion of 
even powers of the Vandermonde determinant into Schur functions 
\cite{Dunne,Scharf} turn out to be especially efficient.
For $\gamma$ being an even integer, representations based on
the permutation group \cite{Samaj2,Tellez} are useful.
At arbitrary integer $\gamma$, the 2D OCP is mappable
onto a discrete 1D fermionic field theory \cite{Samaj3}.
Within this fermionic representation, a symmetry of the model
with respect to a complex transformation of particle coordinates
has been shown to imply a functional relation for the two-body density.
The functional relation is equivalent to an infinite sequence of sum rules
relating the coefficients of the short-distance expansion of the
two-body density.
The generalization of the symmetry to multi-particle densities,
possessing a specific invariant structure, was presented
in Ref. \cite{Samaj4}.
 
The mathematical formulation of the 2D OCP looks at first sight
simpler than the one of the 2D TCP.
The mentioned integrability of the Coulomb gas therefore evokes
the potential possibility of the integrability of the 2D OCP, and 
this is the main subject of the present paper.
The integrability of the 2D jellium is investigated first at arbitrary
coupling $\Gamma$ via an equivalent 2D Euclidean-field theory, 
and then at special values $\Gamma=2*$integer via the equivalent 
1D fermionic model introduced in Ref. \cite{Samaj3}.
As a by-product of the developed formalism, the gauge invariance
of the 2D OCP is proven at the free-fermion point $\Gamma=2$. 

The paper is organized as follows.
 
Section 2 is devoted to the 2D Euclidean-field representation of the
2D OCP.
In Subsection 2.1., we sketch shortly the phenomenological
Debye-H\"uckel calculation of the free energy in order to have
a test formula for functional methods.  
The 2D OCP is mapped onto a 2D Euclidean-field theory
in Subsection 2.2.
A comparison with the sine-Gordon representation of the 2D TCP
is made in Subsection 2.3.
In Subsection 2.4., the ``classical'' integrability of the Euclidean-field 
representation of the 2D OCP is investigated by using a scheme 
proposed by Ghoshal and Zamolodchikov \cite{Ghoshal}.

Section 3 is devoted to a further development of the discrete 1D 
fermionic representation of the 2D OCP at couplings
$\Gamma=2*$integer \cite{Samaj3}.
At these couplings, the partition function of the plasma is shown 
to admit a representation in terms of a linear set of equations.

Section 4 deals with gauge invariance of the bulk 2D OCP at coupling
$\Gamma=2$ which has been proven previously by more standard methods in 
Ref. \cite{Cornu}.
The alternative proof of gauge invariance presented here is related 
to the exact inversion of a class of infinite-dimensional matrices, 
which is of mathematical interest.

A brief recapitulation is given in Section 5.

\renewcommand{\theequation}{2.\arabic{equation}}
\setcounter{equation}{0}

\section{2D FIELD REPRESENTATION}

\subsection{Debye-H\"uckel Calculation}
In the mean-field approximation, the effective electric potential 
$\psi$ at distance $r$ of charge $q$, placed at the origin ${\bf 0}$ 
and surrounded by mobile $q$-charges plus the neutralizing background, 
is given by the 2D Poisson equation
\begin{equation} \label{2.1}
\Delta \psi({\bf r}) = - 2 \pi q \left\{
\delta({\bf r}) + n \left[ {\rm e}^{-\beta q \psi({\bf r})}
-1 \right] \right\}
\end{equation}
The mean-field Boltzmann factor can be linearized for high temperatures.
Eq. (\ref{2.1}) then transforms to
\begin{equation} \label{2.2}
\left( \Delta - \kappa^2 \right) \psi({\bf r})
= - 2\pi q \delta({\bf r})
\end{equation}
where $\kappa$ is the inverse Debye length defined by
$\kappa^2 = 2\pi \Gamma n$.
The solution of (\ref{2.2}) reads $\psi({\bf r}) = q K_0(\kappa r)$,
where $K_0$ is a modified Bessel function.
The excess (i.e., over ideal) free energy per particle, $f_{\rm ex}$,
is expressible in terms of the excess potential energy per particle
\begin{equation} \label{2.3}
u_{\rm ex}(\beta) = \frac{q}{2} \lim_{r\to 0}
\left[ \psi(r) + q \ln r \right]
\end{equation}
as follows
\begin{equation} \label{2.4}
\beta f_{\rm ex} = \int_0^{\beta} {\rm d}\beta'
u_{\rm ex}(\beta')
\end{equation}
Using the short-distance expansion of $K_0$ \cite{Gradshteyn},
\begin{equation} \label{2.5}
K_0(x) = - \ln(x/2) - C + O(x^2\ln x)
\end{equation}
$C$ is the Euler's constant, we arrive at the expression
\begin{equation} \label{2.6}
\beta f_{\rm ex} \sim - \frac{\Gamma}{4} \ln \left( \frac{\kappa^2}{4} \right)
+ \frac{\Gamma}{4} \left( 1 - 2 C \right)
\end{equation} 
valid in the small coupling limit $\Gamma\to 0$.
In what follows, this will be a test formula for functional methods.

An analogous procedure can be applied to the $\Gamma\to 0$
limit of the 2D TCP of $\pm q$ charged particles.
The excess free energy per particle is again obtained 
in the form (\ref{2.6}).

\subsection{Field-Theoretical Representation}
The 2D Coulomb potential (\ref{1.2}) is singular at $r=0$.
This causes mathematical difficulties when representing
interacting Coulomb systems as equivalent field theories.
To avoid this problem, we will consider the Coulomb
potential regularized smoothly at short distances:
\begin{equation} \label{2.7}
v_{\rm reg}(r) = -\ln r - K_0\left( \frac{r}{\epsilon}\right) ,
\quad \quad \epsilon > 0
\end{equation}
In 3D, the analogous regularization has been used 
in Ref. \cite{Brydges}.
Since the Bessel function $K_0(x)$ decays to zero exponentially as
$x\to \infty$, $v_{\rm reg}(r)$ has the large-$r$ asymptotic of
the pure Coulomb potential.
On the other hand, using the short-distance expansion (\ref{2.5})
in (\ref{2.7}), the self-energy is finite
\begin{equation} \label{2.8}
v_{\rm reg}(0) = C - \ln 2 - \frac{1}{2} \ln \epsilon^2
\end{equation}
It is easy to verify that the regularized Coulomb potential 
satisfies the differential equation
\begin{equation} \label{2.9}
\left( \Delta - \epsilon^2 \Delta^2 \right) v_{\rm reg}({\bf r})
= - 2 \pi \delta({\bf r})
\end{equation}
which is the counterpart of the 2D Poisson equation (\ref{1.1}).

We are interested in the bulk thermodynamic properties of the OCP,
defined in the infinite region $\Lambda = R^2$ with the volume
$\vert \Lambda \vert \to \infty$.
For $N$ mobile particles at positions $\{ {\bf r}_j \}_{j=1}^N$
in $\Lambda$, we introduce the microscopic density of particles
${\hat n}$ and of the total charge ${\hat \rho}$ as follows
\begin{equation} \label{2.10}
{\hat n}({\bf r}) = \sum_{j=1}^N \delta({\bf r}-{\bf r}_j) , \quad
{\hat \rho}({\bf r}) = q \sum_{j=1}^N \delta({\bf r}-{\bf r}_j) - q n_b
\end{equation}
Here, $-q n_b$ is the fixed (i.e., $N$-independent) charge density
of the background.
The total interaction energy of the particle-background system
is expressible as
\begin{equation} \label{2.11}
E_N(\{ {\bf r}_j \}) = \frac{1}{2} 
\int_{\Lambda} {\rm d}^2 r \int_{\Lambda} {\rm d}^2 r'
{\hat \rho}({\bf r}) v_{\rm reg}(\vert {\bf r}-{\bf r}' \vert)
{\hat \rho}({\bf r}') - \frac{1}{2} N q^2 v_{\rm reg}(0)  
\end{equation}
We will work in the grand canonical ensemble with the fixed
background \cite{Lieb,Fantoni} and position-dependent 
fugacity $z({\bf r})$ of particles.
The grand partition function $\Xi$ at inverse temperature $\beta$
is defined as the sum over all $N$-particle states
\begin{equation} \label{2.12}
\Xi[z] = \sum_{N=0}^{\infty} \frac{1}{N!} 
\int \prod_{j=1}^N \left[ {\rm d}^2 r_j z({\bf r}_j) \right]
{\rm e}^{-\beta E_N(\{ {\bf r}_j\})}
\end{equation}
The multi-particle densities can be obtained as the functional
derivatives of the generator $\Xi$ with respect to $z({\bf r})$;
after the functional derivatives are done, the homogeneous regime
with the uniform fugacity $z({\bf r}) = z$ is considered.
At the one-particle level, the particle density is given by
\begin{eqnarray}
n & = & \langle {\hat n}({\bf r}) \rangle \nonumber \\
& = & z({\bf r}) \frac{1}{\Xi} 
\frac{\delta \Xi}{\delta z({\bf r})}\bigg\vert_{\rm uniform}
\label{2.13}
\end{eqnarray}
At the two-particle level, one introduces the two-body density
\begin{eqnarray}
n_2({\bf r},{\bf r}') & = & 
\langle {\hat n}({\bf r}) {\hat n}({\bf r}') \rangle
- n \delta({\bf r}-{\bf r}') \nonumber \\
& = & z({\bf r}) z({\bf r}') \frac{1}{\Xi}
\frac{\delta^2 \Xi}{\delta z({\bf r}) \delta z({\bf r}')}
\bigg\vert_{\rm uniform} \label{2.14}
\end{eqnarray}

The grand partition function (\ref{2.12}) can be expressed in terms
of a 2D Euclidean-field theory.
We start by the standard procedure (see, e.g., Ref. \cite{Minnhagen})
and substitute the representation (\ref{2.11}) of $E_N$
in the Boltzmann factor $\exp(-\beta E_N)$.
The self-energy term renormalizes the fugacity,
$z({\bf r}) \to {\bar z}({\bf r}) = 
z({\bf r}) \exp[\Gamma v_{\rm reg}(0)/2]$.
According to relation (\ref{2.9}), $-(\Delta-\epsilon^2\Delta^2)/(2\pi)$
is the inverse operator of $v_{\rm reg}$.
The bilinear term in $\exp(-\beta E_N)$ can thus be linearized by
applying the Hubbard-Stratonovich transformation
\begin{eqnarray}
& & \exp\left[ -\frac{\beta}{2} \int {\rm d}^2 r \int {\rm d}^2 r'
{\hat \rho}({\bf r}) v_{\rm reg}(\vert {\bf r}-{\bf r}'\vert)
{\hat \rho}({\bf r}') \right] \nonumber \\
& & \quad \quad = \frac{\int {\cal D}\phi \exp\left\{ \int {\rm d}^2 r
\left[ \frac{1}{2} \phi (\Delta-\epsilon^2\Delta^2) \phi
+{\rm i} \sqrt{2\pi\beta} \phi {\hat \rho} \right] \right\}}{\int
{\cal D}\phi \exp\left[ \int {\rm d}^2 r \frac{1}{2}
\phi (\Delta - \epsilon^2 \Delta^2) \phi \right]} \label{2.15}
\end{eqnarray}
Here, $\phi({\bf r})$ is a real scalar field with all derivatives
vanishing at infinity and $\int {\cal D}\phi$ denotes the 
functional integration over this field.
The terms $\phi \Delta \phi$ and $\phi \Delta^2 \phi$ can be turned
into $-\vert \nabla\phi\vert^2$ and $(\Delta\phi)^2$, respectively,
after performing integrations by parts with vanishing contributions
from infinity.
Inserting ${\hat\rho}$ from (\ref{2.10}), particle coordinates
in (\ref{2.12}) become decoupled from each other and one can sum
over $N$, with the result
\begin{equation} \label{2.16}
\Xi = \int \frac{{\cal D}\phi}{D} 
\exp\left\{ - \int {\rm d}^2 r \left[ 
\frac{1}{2}\vert \nabla \phi \vert^2 
+ \frac{1}{2} \epsilon^2 (\Delta \phi)^2
- {\bar z}({\bf r}) {\rm e}^{{\rm i}\sqrt{2\pi\Gamma}\phi}
+ {\rm i} \sqrt{2\pi\Gamma} n_b \phi \right] \right\}
\end{equation} 
where
\begin{equation} \label{2.17}
D = \int {\cal D}\phi \exp\left\{ - \int {\rm d}^2 r \left[ 
\frac{1}{2}\vert \nabla \phi \vert^2 
+ \frac{1}{2} \epsilon^2 (\Delta \phi)^2 \right] \right\}
\end{equation}
is the normalization constant.
In the homogeneous regime ${\bar z}({\bf r}) = {\bar z}$,
the uniform shift in $\phi$
\begin{equation} \label{2.18}
\phi({\bf r}) \rightarrow \phi({\bf r}) +
\frac{\ln ({\bar z}/n_b)}{{\rm i}\sqrt{2\pi\Gamma}}
\end{equation}
factorizes out the $z$-dependence of $\Xi$,
\begin{eqnarray}
\Xi & = & \exp\left[ \vert \Lambda \vert 
\ln\left(\frac{\bar z}{n_b} \right)
+ \vert \Lambda \vert n_b \right] \int \frac{{\cal D}\phi}{D} 
\label{2.19} \\
& \times & \exp\left\{ - \int {\rm d}^2 r \left[ 
\frac{1}{2}\vert \nabla \phi \vert^2 
+ \frac{1}{2} \epsilon^2 (\Delta \phi)^2
+ n_b \left( - {\rm e}^{{\rm i}\sqrt{2\pi\Gamma}\phi} + 1
+ {\rm i} \sqrt{2\pi\Gamma} \phi \right) \right] \right\}
\nonumber
\end{eqnarray}
The particle density $n$ is yielded by the homogeneous
analogue of Eq. (\ref{2.13}) as follows
\begin{equation} \label{2.20}
n = z \frac{\partial}{\partial z} \left(
\frac{\ln \Xi}{\vert \Lambda\vert} \right) = n_b
\end{equation}
This means that from the grand partition sum (\ref{2.12}) only the term 
with the strict system neutrality survives, in the spirit of
Ref. \cite{Lieb}.
The density-fugacity relationship is trivial, namely the density
does not depend on the fugacity.
This enables us to pass to the canonical ensemble via the
Legendre transformation
\begin{equation} \label{2.21}
-\beta F(n) = \ln \Xi - N \ln z
\end{equation}
where $F$ is the Helmholtz free energy and $N= n \vert\Lambda\vert$.
The excess free energy, related to $F$ as follows
$-\beta F_{\rm ex} = -\beta F + N\ln n - N$, 
then reads
\begin{equation} \label{2.22}
-\beta F_{\rm ex}(n) = \frac{N \Gamma}{2} 
\left( C - \ln 2 - \frac{1}{2} \ln \epsilon^2 \right)
+ \ln R
\end{equation}
where we have substituted the explicit form of the self-energy
(\ref{2.8}) and grouped the field part into the quantity $R$ defined by
\begin{equation} \label{2.23}
R = \frac{
\int {\cal D}\phi \exp\left\{ - \int {\rm d}^2 r \left[ 
\frac{1}{2}\vert \nabla \phi \vert^2 
+ \frac{1}{2} \epsilon^2 (\Delta \phi)^2
+ n \left( - {\rm e}^{{\rm i}\sqrt{2\pi\Gamma}\phi} + 1
+ {\rm i} \sqrt{2\pi\Gamma} \phi \right) \right] \right\}
}{\int {\cal D}\phi \exp\left\{ - \int {\rm d}^2 r \left[ 
\frac{1}{2}\vert \nabla \phi \vert^2 
+ \frac{1}{2} \epsilon^2 (\Delta \phi)^2 \right] \right\}}
\end{equation}

For the pure Coulomb interaction $(\epsilon = 0)$, the field
representation of the free energy similar to the one described
by Eqs. (\ref{2.22}) and (\ref{2.23}) was established directly 
in the canonical format by Brilliantov \cite{Brilliantov}.
The problem of the divergent self-energy was incorrectly
ignored there, although this one enters into the final formulae.
In what follows, we aim at deriving the small-$\Gamma$ expansion
of $\ln R$ in (\ref{2.22}) in order to show that, in the limit
$\epsilon\to 0$, the divergent self-energy term $\propto \ln\epsilon^2$
is canceled and the Debye-H\"uckel result (\ref{2.6}) is reproduced
correctly.
For small $\Gamma$, we expand the exponential
\begin{equation} \label{2.24}
{\rm e}^{{\rm i}\sqrt{2\pi\Gamma}\phi} \sim
1 + {\rm i}\sqrt{2\pi\Gamma}\phi 
- \frac{1}{2}(2\pi\Gamma) \phi^2
\end{equation}
$R$ then becomes equal to
\begin{equation} \label{2.25}
R = \frac{
\int {\cal D}\phi \exp\left\{ - \int {\rm d}^2 r \left[ 
\frac{1}{2}\vert \nabla \phi \vert^2 
+ \frac{1}{2} \epsilon^2 (\Delta \phi)^2
+ \frac{1}{2} \kappa^2 \phi^2 \right] \right\}
}{\int {\cal D}\phi \exp\left\{ - \int {\rm d}^2 r \left[ 
\frac{1}{2}\vert \nabla \phi \vert^2 
+ \frac{1}{2} \epsilon^2 (\Delta \phi)^2 \right] \right\}}
\end{equation} 
The Gaussian functional integrals can be diagonalized in
the Fourier ${\bf k}$-space, with the result
\begin{equation} \label{2.26}
R = \prod_k \left( 
\frac{k^2+\epsilon^2 k^4}{\kappa^2+k^2+\epsilon^2 k^4} \right)^{1/2}
= \exp\left\{ \frac{1}{2} \vert \Lambda \vert 
\int \frac{{\rm d}^2 k}{(2\pi)^2}
\ln \left( \frac{k^2+\epsilon^2 k^4}{\kappa^2+k^2+\epsilon^2 k^4} \right)
\right\}
\end{equation}
The integration over ${\bf k}$ can be carried out explicitly
and one finds
\begin{equation} \label{2.27}
\ln R = \frac{\vert \Lambda \vert}{8\pi} \left(
t_+ \ln t_+ + t_- \ln t_- + \frac{1}{\epsilon^2} \ln \epsilon^2 \right) 
\end{equation}
where
\begin{equation} \label{2.28}
t_{\pm} = \frac{1 \pm \sqrt{1-4\epsilon^2\kappa^2}}{2\epsilon^2}
\end{equation}
In the $\epsilon\to 0$ limit,
\begin{equation} \label{2.29}
t_+ = \frac{1}{\epsilon^2} - \kappa^2 + O(\epsilon^2), \quad
t_- = \kappa^2 + O(\epsilon^2)
\end{equation}
Consequently,
\begin{equation} \label{2.30}
\ln R \sim \frac{N\Gamma}{4} 
\left( \ln \epsilon^2 + \ln \kappa^2 -1 \right) \quad
{\rm as}\ \epsilon\to 0
\end{equation}
Inserting this into (\ref{2.22}), the singular $\ln \epsilon^2$ term
disappears and one recovers the Debye-H\"uckel result (\ref{2.6}).

The two-body density can be obtained from the field representation 
(\ref{2.16}) of $\Xi[z]$ using formula (\ref{2.14}).
The uniform shift in the $\phi$-field, relation (\ref{2.18}), 
then leads to
\begin{equation} \label{2.31}
\frac{n_2({\bf r},{\bf r}')}{n({\bf r}) n({\bf r}')} = \frac{
\langle {\rm e}^{{\rm i}\sqrt{2\pi\Gamma}\phi({\bf r})}
{\rm e}^{{\rm i}\sqrt{2\pi\Gamma}\phi({\bf r}')} \rangle}
{\langle {\rm e}^{{\rm i}\sqrt{2\pi\Gamma}\phi({\bf r})} \rangle
\langle {\rm e}^{{\rm i}\sqrt{2\pi\Gamma}\phi({\bf r}')} \rangle}
\end{equation}
where the averages $\langle \cdots \rangle$
are taken with the field action
\begin{equation} \label{2.32}
S[\phi] = \int {\rm d}^2 r \left[
\frac{1}{2} \vert \nabla\phi \vert^2
+ \frac{1}{2} \epsilon^2 (\Delta\phi)^2
+ n \left( - {\rm e}^{{\rm i}\sqrt{2\pi\Gamma}\phi} + 1
+ {\rm i} \sqrt{2\pi\Gamma} \phi \right) \right]
\end{equation}
Note that because the self-energy does not enter explicitly into 
(\ref{2.31}), one can put $\epsilon=0$ in the action (\ref{2.32})
(the consequent singularities in the numerator and the denominator
must be precisely canceled with one another), and consider
\begin{equation} \label{2.33}
S[\phi] = \int {\rm d}^2 r \left[
\frac{1}{2} \vert \nabla\phi \vert^2
+ n \left( - {\rm e}^{{\rm i}\sqrt{2\pi\Gamma}\phi} + 1
+ {\rm i} \sqrt{2\pi\Gamma} \phi \right) \right]
\end{equation}

In the Debye-H\"uckel $\Gamma\to 0$ limit, the expansion of the
exponential according to Eq. (\ref{2.24}) transforms the action
(\ref{2.33}) to
\begin{equation} \label{2.34}
S_{\rm DH} = \int {\rm d}^2 r \left[
\frac{1}{2} \vert \nabla\phi \vert^2
+ \frac{1}{2} \kappa^2 \phi^2 \right]
\end{equation}
Since $\langle \phi \rangle = 0$ with this action, the Wick's 
theorem for Gaussian integrals implies
\begin{eqnarray}
\langle {\rm e}^{{\rm i}\sqrt{2\pi\Gamma}\phi({\bf r})} \rangle
& = & \exp\left\{ -\pi\Gamma \langle \phi^2({\bf r}) \rangle \right\}
\label{2.35} \\
\langle {\rm e}^{{\rm i}\sqrt{2\pi\Gamma}[\phi({\bf r})+\phi({\bf r}')]} 
\rangle & = & 
\exp\left\{ -\pi\Gamma \langle [\phi({\bf r})+\phi({\bf r}')]^2 
\rangle \right\} \label{2.36} 
\end{eqnarray}
Consequently,
\begin{equation} \label{2.37}
\frac{n_2({\bf r},{\bf r}')}{n^2} = \exp\left\{
-2\pi\Gamma \langle \phi({\bf r}) \phi({\bf r}') \rangle \right\}
\end{equation}
For the quadratic action (\ref{2.34}), the correlator 
$\langle \phi({\bf r}) \phi({\bf r}') \rangle$
is equal to the inverse matrix element of the operator $-\nabla^2+\kappa^2$,
\begin{equation} \label{2.38}
\langle \phi({\bf r}) \phi({\bf r}') \rangle =
\frac{1}{2\pi} K_0(\kappa\vert {\bf r}-{\bf r}'\vert )
\end{equation} 
Thence
\begin{equation} \label{2.39}
\frac{n_2({\bf r},{\bf r}')}{n^2} = 
{\rm e}^{-\Gamma K_0(\kappa\vert {\bf r}-{\bf r}'\vert )}
\sim 1 - \Gamma K_0(\kappa\vert {\bf r}-{\bf r}'\vert )
\end{equation}
in agreement with the standard Debye-H\"uckel calculation 
(see, e.g., Ref. \cite{Kalinay1}).

\subsection{Comparison with the 2D TCP}
It is well known that the 2D TCP of $\pm q$ charges is equivalent
to the sine-Gordon theory \cite{Minnhagen}.
Namely, the grand partition function $\Xi$ is expressible as
\begin{equation} \label{2.40}
\Xi(z) = \frac{\int {\cal D}\phi \exp\left( -S(z) \right)}{\int 
{\cal D}\phi \exp\left( -S(0) \right)}
\end{equation}
where $S$ is the 2D Euclidean sine-Gordon action
\begin{equation} \label{2.41}
S(z) = \int {\rm d}^2 r \left[ \frac{1}{2}\vert \nabla \phi\vert^2
- 2 {\bar z} \cos\left( \sqrt{2\pi\Gamma} \phi \right) \right]
\end{equation}
As before, ${\bar z}$ is the particle fugacity renormalized by the
(divergent) self-energy term, ${\bar z} = z\exp[\beta v(0)/2]$.
In contrast to the previous case of the OCP, the self-energy
$v(0)$ cannot be eliminated from the functional integration
through a uniform shift in $\phi$.
To give a precise meaning to ${\bar z}$ one has to fix the
normalization of the coupled cos-field.
In the Coulomb gas, the behavior of the two-body density
for oppositely charged particles is dominated at short distance
by the Boltzmann factor of the Coulomb potential,
\begin{equation} \label{2.42}
n_{+-}({\bf r},{\bf r}') \sim z^2 \vert {\bf r}-{\bf r}'\vert^{-\Gamma}
\quad {\rm as}\ \vert {\bf r}-{\bf r}' \vert \to 0
\end{equation} 
which is the crucial supplement of the mapping (\ref{2.40}) and
(\ref{2.41}).
Under this conformal short-distance normalization, the divergent 
self-energy factor disappears from statistical relations 
established within the sine-Gordon formulation.

One can document the above scheme in the Debye-H\"uckel limit
$\Gamma\to 0$, when 
$\cos(\sqrt{2\pi\Gamma}\phi) \sim 1 - (2\pi\Gamma)\phi^2/2$ and
\begin{equation} \label{2.43}
S_{\rm DH} = \int {\rm d}^2 r \left[ \frac{1}{2} \vert \nabla \phi \vert^2
+ 2\pi \Gamma {\bar z} \phi^2 \right] - 2 {\bar z} \vert \Lambda \vert
\end{equation} 
The total particle density and the two-body density of the
opposite charges are given by
\begin{eqnarray}
n & = & 2 {\bar z} 
\langle {\rm e}^{{\rm i}\sqrt{2\pi\Gamma}\phi({\bf r})} \rangle
\label{2.44} \\
n_{+-}({\bf r},{\bf r}') & = & {\bar z}^2
\langle {\rm e}^{{\rm i}\sqrt{2\pi\Gamma}\phi({\bf r})}
{\rm e}^{-{\rm i}\sqrt{2\pi\Gamma}\phi({\bf r}')} \rangle
\label{2.45}
\end{eqnarray}
respectively; the averages are taken with the action $S_{\rm DH}$.
The Wick's theorem for Gaussian integrals gives
\begin{eqnarray}
\langle {\rm e}^{{\rm i}\sqrt{2\pi\Gamma}\phi({\bf r})} \rangle
& = & \exp\left\{ -\pi\Gamma \langle \phi^2({\bf r}) \rangle \right\}
\label{2.46} \\
\langle {\rm e}^{{\rm i}\sqrt{2\pi\Gamma}[\phi({\bf r})-\phi({\bf r}')]} 
\rangle & = & 
\exp\left\{ -\pi\Gamma \langle [\phi({\bf r})-\phi({\bf r}')]^2 
\rangle \right\} \label{2.47} 
\end{eqnarray}
Simultaneously,
\begin{equation} \label{2.48}
\langle \phi({\bf r}) \phi({\bf r}') \rangle = \frac{1}{2\pi}
K_0\left( \sqrt{4\pi\Gamma{\bar z}} \vert {\bf r}-{\bf r}' \vert\right)
\end{equation}
with the short-distance asymptotic
\begin{equation} \label{2.49}
\langle \phi({\bf r}) \phi({\bf r}') \rangle \sim
\frac{1}{2\pi} \left[ - \ln\left( 
\sqrt{\pi\Gamma{\bar z}} \vert {\bf r}-{\bf r}' \vert\right) - C \right]
\quad {\rm as}\ \vert {\bf r}-{\bf r}' \vert \to 0 
\end{equation}
Combining Eqs. (\ref{2.44})-(\ref{2.49}) with the short-distance 
normalization (\ref{2.42}), one arrives at the explicit density-fugacity
relation
\begin{equation} \label{2.50}
\frac{n^{1-\Gamma/4}}{2 z} = \left( \frac{\pi\Gamma}{2} \right)^{\Gamma/4}
\exp \left( \frac{\Gamma C}{2} \right)
\end{equation}
valid in the leading $\Gamma$-order.
This relation does not contain the self-energy $v(0)$; the latter
was coupled only to the $\Gamma^2$ power.
Having the explicit $n-z$ relation it is straightforward to pass
to the canonical format.
The Debye-H\"uckel relation (\ref{2.6}) is reproduced again.

\subsection{Classical Non-Integrability}
The field action of the 2D OCP (\ref{2.33}) belongs to a more general
class of actions possessing the local form
\begin{equation} \label{2.51}
S[\phi] = \int {\rm d}^2 r \left[
\frac{1}{2} (\partial_x\phi)^2 + \frac{1}{2} (\partial_y\phi)^2
+ 4 V(\phi) \right]
\end{equation}
where the factor 4 in the potential term appears for
notation convenience.
The exact solvability of a theory depends on the particular form of 
the potential $V$, which is in our case
\begin{equation} \label{2.52}
V(\phi) = \frac{n}{4} \left(
- {\rm e}^{{\rm i}\sqrt{2\pi\Gamma}\phi} + 1
+ {\rm i}\sqrt{2\pi\Gamma} \phi \right)
\end{equation}
It is well known \cite{Zinn} that the 2D Euclidean field theory
(\ref{2.51}), defined in the space of points ${\bf r}=(x,y)$,
is the imaginary-time $(y={\rm i}t)$ continuation of the equivalent
real-time 1+1 dimensional quantum field theory with the action
\begin{equation} \label{2.53}
S[\phi] = \int_{-\infty}^{\infty} {\rm d}t 
\int_{-\infty}^{\infty} {\rm d}x \left[
\frac{1}{2} (\partial_t\phi)^2 - \frac{1}{2} (\partial_x\phi)^2
- 4 V(\phi) \right]
\end{equation}
This action is dominated by fields satisfying the ``classical''
equation of motion $\delta S = 0$.
In terms of light-cone coordinates defined by
$\partial_{\pm} = (\partial_t\pm\partial_x)/2$,
the equation of motion reads
\begin{equation} \label{2.54}
\partial_+ \partial_-  \phi = - V'(\phi)
\end{equation}
In the case of our potential (\ref{2.52}), rescaling appropriately
the $\phi$-field, this equation is nothing but the $1+1$
analogue of the usual mean-field Poisson-Boltzmann equation
for the OCP.
In general, the integrability of a field theory is associated with
existence of an infinite sequence of conserved quantities 
(integrals of motion or ``charges'').
In what follows, we use a general scheme \cite{Ghoshal} 
to find out whether or not there exists an infinite sequence of 
conserved quantities for our field theory with the potential $V$ 
given by (\ref{2.52}), in the classical limit, i.e., when 
the field $\phi$ is governed by the equation of motion (\ref{2.54}).
The scheme represents a unique way of determinig integrability 
properties of the given field theory.

Existence of a conserved charge is associated with the appearance 
of a pair of ``conjugate'' local field densities $(T,\theta)$, 
with zero boundary conditions at $x\to\pm\infty$, such that
\begin{equation} \label{2.55}
\partial_- T = \partial_+ \theta
\end{equation}
In terms of $x$ and $t$ variables this is equivalent to
$\partial_t (T-\theta) = \partial_x(T+\theta)$.
The integration over $x$ results in
\begin{equation} \label{2.56}
\frac{\partial}{\partial t} \left[ \int_{-\infty}^{\infty}
{\rm d}x (T-\theta) \right] = (T+\theta) \big\vert_{-\infty}^{\infty}
= 0
\end{equation}
and thence the charge $\int {\rm d}x (T-\theta)$ is conserved.
We look for $(T,\theta)$ as polynomial functions in derivatives
$\partial_+\phi, \partial_+^2\phi$, etc.
The notation $T_s$ ($\theta_s$) will be used for $T$ ($\theta$)
with just $s$ $\partial_+$-derivatives.
Because of the specific form of the equation of motion (\ref{2.54}),
only $T_{s+1}$ and $\theta_{s-1}$ can create the conjugate couple,
$\partial_- T_{s+1} = \partial_+ \theta_{s-1}$.
The conserved charge is then
\begin{equation} \label{2.57}
Q_s = \int {\rm d}x (T_{s+1}-\theta_{s-1}) \quad
s = 1, 2, \ldots
\end{equation}
Note that the total $\partial_+$ derivatives can be dropped from
$T$ because the consequent difference 
$\partial_+ - \partial_- = \partial_x$ produces vanishing
boundary contributions to the conserved charge.
Similarly, let ${\tilde T}_{s-1}$ and ${\tilde \theta}_{s+1}$
be the conjugate polynomials of given orders in 
$\partial_-$-derivatives such that
$\partial_- {\tilde T}_{s-1} = \partial_+ {\tilde \theta}_{s+1}$.
Then, the charge
\begin{equation} \label{2.58}
{\tilde Q}_s = \int {\rm d}x ({\tilde\theta}_{s+1}-{\tilde T}_{s-1}) 
\quad s = 1, 2, \ldots
\end{equation}
is conserved.
As before, the total $\partial_-$ derivatives can be dropped 
from ${\tilde\theta}$.

At $s=1$, writing $T_2 = (\partial_+\phi)^2$ one has
\begin{equation} \label{2.59}
\partial_- T_2 = 2 (\partial_+\phi)(\partial_+\partial_-\phi)
= -2 (\partial_+\phi) V'(\phi) = \partial_+ \left[ -2 V(\phi) \right]
\end{equation}
so that $\theta_0=-2V$ and 
\begin{equation} \label{2.60}
Q_1 = \int {\rm d}x \left[ (\partial_+\phi)^2 + 2 V \right]
\end{equation}
Analogously, ${\tilde\theta}_2 = (\partial_-\phi)^2$, 
${\tilde T}_0 = -2V$ and
\begin{equation} \label{2.61}
{\tilde Q}_1 = \int {\rm d}x \left[ (\partial_-\phi)^2 + 2 V \right]
\end{equation}
$Q_1+{\tilde Q}_1$ and $Q_1-{\tilde Q}_1$ are energy and momentum,
respectively, and these two quantities are always conserved for
any potential $V$.
There is no solution for conjugate polynomials producing conserved charges
at $s$ being an even integer.
For $s=3$, there exist conjugate polynomials
\begin{eqnarray} 
T_4 & = & \left( \frac{b}{2} \right)^2 (\partial_+\phi)^4
+ (\partial_+^2 \phi)^2 \label{2.62} \\
\theta_2 & = & - \left( \partial_+\phi \right)^2 V''(\phi)
\label{2.63}
\end{eqnarray}
and the corresponding $({\tilde\theta}_4,{\tilde T}_2)$,
provided that the potential satisfies the differential equation
$V'''=b^2 V'$.
This equation is fulfilled either for the trivial free field theory
$(b=0)$ or for the potential
\begin{equation} \label{2.64}
V(\phi) = A {\rm e}^{b\phi} + B {\rm e}^{-b\phi},
\quad b\ne 0
\end{equation}
When the constants $A$ and $B$ are nonzero and equal to one another,
one recognizes the sinh-Gordon ($b$ real) or sine-Gordon ($b$ imaginary) 
models.
At $s=5$, there exist conjugate polynomials and the corresponding
conserved charges if the potential is either of the previous
form (\ref{2.64}) or of the form
\begin{equation} \label{2.65}
V(\phi) = A {\rm e}^{b\phi} + B {\rm e}^{-(b/2)\phi},
\quad b\ne 0
\end{equation}
This integrable field theory is known as the Bullough-Dodd model
\cite{Dodd} and for imaginary $b$ it corresponds to the $1:2$
charge-asymmetric Coulomb gas \cite{Samaj5}.
The models with potentials (\ref{2.64}) and (\ref{2.65}) are the only two
one-component-field members of the integrable affine Toda field theories,
based on the Dynkin-diagram classification of simple Lie groups.
Programming the whole scheme in the symbolic language {\it Reduce}, 
we were able to proceed up to the relatively high $s=15$ order.
Except for the repeated appearance of the two potentials (\ref{2.64}) 
and (\ref{2.65}), we did not find any other solution for the potential
leading to conserved charges.
We do not anticipate a sudden appearance of an additional
potential producing conserved charges for $s>15$. 
If it is so the 2D OCP, characterized by the field potential (\ref{2.52}),
is not classically integrable.

The conjectured classical non-integrability does not exclude 
the complete ``quantum'' (all realizations
of the $\phi$-field are considered) integrability of the model.
At specific values of the coupling constant $\Gamma$, 
quantum fluctuations of the field around its classical 
saddle-point value can make the plasma
integrable, as it is at the free-fermion point $\Gamma=2$. 

\renewcommand{\theequation}{3.\arabic{equation}}
\setcounter{equation}{0}

\section{LINEAR FERMIONIC REPRESENTATION}
We now consider the 2D OCP, confined to a domain $\Lambda$,
directly in the canonical ensemble.
$N$ pointlike $q$-charges are embedded in a spatially homogeneous
background of charge density $\rho_b = -q n_b$.
If the system-neutrality condition is imposed it holds 
$n_b=N/\vert \Lambda \vert$.
The background produces the one-body electric potential
$v_b({\bf r}) = \rho_b \int_{\Lambda} {\rm d}^2 r'
v(\vert {\bf r}-{\bf r}'\vert)$ which satisfies the Poisson equation
\begin{equation} \label{3.1}
\Delta v_b({\bf r}) = 2\pi q n_b , \quad {\bf r}\in \Lambda
\end{equation}
Since, written in the complex $(z,{\bar z})$-coordinates,
\begin{equation} \label{3.2}
\Delta = 4 \partial_z \partial_{\bar z}
\end{equation}
for a circularly symmetric background one has
\begin{equation} \label{3.3}
v_{\rm circ} = {\rm const} + \frac{\pi q n_b}{2} z {\bar z}
\end{equation}
The deformation of the circular boundary $\partial\Lambda$
or the presence of some charge densities outside of $\Lambda$
generates an additional gauge potential $v_{\rm gauge}$
such that $\Delta v_{\rm gauge}(z,{\bar z}) = 0$ for
$z\in \Lambda$.
With regard to (\ref{3.2}), the general solution of this equation
reads
\begin{equation} \label{3.4}
v_{\rm gauge}(z,{\bar z}) = A_0 + 
\sum_{s=1}^{\infty} \left( A_s z^s + B_s {\bar z}^s \right)
\end{equation}
Since Coulomb potentials are real, physical situations
correspond to the choice $A_s={\bar B}_s$ for all $s$.
The particular case of a quadrupolar potential
$v_{\rm gauge} = A(z^2 + {\bar z}^2)$ results from
the deformation of the disk into an ellipse
\cite{Francesco,Forrester3}.

The potential energy of $N$ particles at positions 
$\{ z_i\in \Lambda\}$ plus the background is
\begin{equation} \label{3.5}
E = E_0 + q \sum_i v_b(z_i,{\bar z}_i) - 
q^2 \sum_{i<j} \ln \vert z_i-z_j \vert 
\end{equation}
The background-background interaction constant $E_0$ does
not influence the particle densities and so it can be omitted.
The partition function at inverse temperature $\beta$ reads
\begin{equation} \label{3.6}
Z_N = \frac{1}{N!} \int_{\Lambda} \prod_{i=1}^N 
\left[ {\rm d}^2 z_i w(z_i,{\bar z}_i) \right]
\prod_{i<j} \vert z_i - z_j \vert^{\Gamma}
\end{equation}
where $\Gamma=\beta q^2$ and the one-body Boltzmann factor
$w(z,{\bar z}) = \exp[-\beta q v_b(z,{\bar z})]$.
The multi-particle densities can be obtained in the standard
way [see relations (\ref{2.13}) and (\ref{2.14})]:
\begin{eqnarray}
n(z,{\bar z}) & = & w(z,{\bar z}) \frac{1}{Z_N} 
\frac{\delta Z_N}{\delta w(z,{\bar z})} \label{3.7} \\
n_2(z_1,{\bar z}_1\vert z_2,{\bar z}_2) & = & 
w(z_1,{\bar z}_1) w(z_2,{\bar z}_2) \frac{1}{Z_N} 
\frac{\delta^2 Z_N}{\delta w(z_1,{\bar z}_1) w(z_2,{\bar z}_2)} 
\label{3.8}
\end{eqnarray} 
etc.
We will study the special case of the plasma with a ``soft wall''
\cite{Jancovici6} when particles are confined by the background itself.
In particular, for a fixed value of the particle number $N$
one makes the radius of the homogeneous circular background infinite,
$\vert \Lambda\vert \to \infty$.
The particles will gather in a circular region of area
$N/n_b$.
In the limit $N\to\infty$, the soft-wall edge of this region
will have the structure different from the one of the usual 
hard-wall problem, but the particle densities deep in the interior
($\vert z\vert$ finite) will not be touched by the soft wall
and will attain the bulk values.
Also the free energy will be modified only by a surface term.
Within the soft-wall $\vert \Lambda\vert \to \infty$ formulation
of the problem, we will consider two cases:

(i) the circularly symmetric homogeneous background, Eq. (\ref{3.3}), 
with
\begin{equation} \label{3.9}
w(z,{\bar z}) = \frac{1}{\pi} \exp \left( - \frac{\Gamma}{2} \pi n_b 
z{\bar z} \right)
\end{equation}

(ii) the general homogeneous background with the gauge component 
(\ref{3.4}), given by
\begin{equation} \label{3.10}
w(z,{\bar z}) = \exp \left[ - \frac{\Gamma}{2} \pi n_b z{\bar z} 
- \sum_{s=1}^{\infty} (a_s z^s + b_s {\bar z}^s) \right]
\end{equation}
where $a_s=\beta q A_s$ and $b_s=\beta q B_s$.
As was already mentioned above, real physical situations
require $a_s={\bar b}_s$ for all $s$ and the values of gauge parameters 
must be such that the multiple integral determining the partition function 
(\ref{3.6}) does not diverge.

For the coupling $\Gamma = 2\gamma$ ($\gamma$ an integer),
it has been shown in Ref. \cite{Samaj3} that the partition
function (\ref{3.6}) can be expressed as the integral over
two sets of Grassmann variables $\{ \xi_i^{(\alpha)},\psi_i^{(\alpha)} \}$ 
each with $\gamma$ components $(\alpha=1,\ldots,\gamma)$, 
defined on a discrete chain of $N$ sites $i=0,1,\ldots,N-1$ 
and satisfying the ordinary anticommuting algebra \cite{Berezin}, 
as follows:
\begin{eqnarray}
Z_N(\gamma) & = & \int {\cal D}\psi {\cal D}\xi~
{\rm e}^{S(\xi,\psi)} \label{3.11} \\
S(\xi,\psi) & = & \sum_{i,j=0}^{\gamma(N-1)}
\Xi_i w_{ij} \Psi_j \label{3.12}
\end{eqnarray}
Here, ${\cal D}\psi {\cal D}\xi = \prod_{i=0}^{N-1}
{\rm d}\psi_i^{(\gamma)}\ldots {\rm d}\psi_i^{(1)}
{\rm d}\xi_i^{(\gamma)}\ldots {\rm d}\xi_i^{(1)}$
and $S$ involves pair interactions of ``composite'' operators
\begin{equation} \label{3.13}
\Xi_i = 
\sum_{i_1,\ldots,i_{\gamma}=0\atop (i_1+\ldots+i_{\gamma}=i)}^{N-1}
\xi_{i_1}^{(1)}\ldots \xi_{i_{\gamma}}^{(\gamma)}, \quad \quad
\Psi_j = 
\sum_{j_1,\ldots,j_{\gamma}=0\atop (j_1+\ldots+j_{\gamma}=j)}^{N-1}
\psi_{j_1}^{(1)}\ldots \psi_{j_{\gamma}}^{(\gamma)}
\end{equation}
i.e., the products of all $\gamma$ anticommuting-field components,
belonging to either $\xi$- or $\psi$-set, 
with the fixed sum of site indices.
The interaction strength is given by
\begin{equation} \label{3.14}
w_{ij} = \int_{\Lambda} {\rm d}^2 z~ w(z,{\bar z}) z^i {\bar z}^j;
\quad \quad i,j = 0,1,\ldots,\gamma(N-1)
\end{equation}
Using the notation $\langle\cdots\rangle = \int {\cal D}\psi
{\cal D}\xi {\rm e}^S \cdots / Z_N(\gamma)$ for an averaging
over the anticommuting variables, the particle density (\ref{3.7})
and the two-body density (\ref{3.8}) are expressible in the
fermionic format as follows
\begin{eqnarray}
n(z,{\bar z}) & = & w(z,{\bar z}) \sum_{i,j=0}^{\gamma(N-1)}
\langle \Xi_i \Psi_j \rangle z^i {\bar z}^j \label{3.15} \\
n_2(z_1,{\bar z}_1\vert z_2,{\bar z}_2) & = & 
w(z_1,{\bar z}_1) w(z_2,{\bar z}_2) \nonumber \\
& & \times
\sum_{i_1,j_1,i_2,j_2=0}^{\gamma(N-1)}
\langle \Xi_{i_1} \Psi_{j_1} \Xi_{i_2} \Psi_{j_2} \rangle 
z_1^{i_1} {\bar z}^{j_1} z_2^{i_2} {\bar z}_2^{j_2} \label{3.16}
\end{eqnarray} 
respectively.

The exact solvability of the 2D OCP at $\Gamma=2$ $(\gamma=1)$
is due to the bilinear form of 
$S=\sum_{i,j=0}^{\gamma(N-1)}\xi_i w_{ij} \psi_j$.
Thus,
\begin{equation} \label{3.17}
Z_N(\gamma=1) = {\rm Det} (w_{ij}) \vert_{i,j=0}^{N-1}
\end{equation}
The two-correlators determining the particle density (\ref{3.15})
are equal to the inverse elements of the 
$N\times N$ {\bf w}-matrix (\ref{3.14}),
\begin{equation} \label{3.18}
\langle \xi_i \psi_j \rangle = w_{ji}^{-1}
\end{equation}
The Wick's theorem applied to the four-correlators in (\ref{3.16})
implies
\begin{eqnarray}
\langle \xi_{i_1} \psi_{j_1} \xi_{i_2} \psi_{j_2} \rangle
& = & \langle \xi_{i_1} \psi_{j_1} \rangle
\langle \xi_{i_2} \psi_{j_2} \rangle -
\langle \xi_{i_1} \psi_{j_2} \rangle
\langle \xi_{i_2} \psi_{j_1} \rangle \nonumber \\
& = & w_{j_1i_1}^{-1} w_{j_2i_2}^{-1}-
w_{j_2i_1}^{-1} w_{j_1i_2}^{-1} \label{3.19}  
\end{eqnarray}

In the case of a circularly symmetric plasma with
$w({\bf r}) = w(r)$ and $\Lambda = \{ r\le R\}$
[like the one of interest, defined by Eq. (\ref{3.9})
and $R\to\infty$], the interaction matrix ${\bf w}$
with elements (\ref{3.14}) becomes diagonal:
\begin{equation} \label{3.20}
w_{ij} = w_i \delta_{ij}, \quad
w_i = 2 \pi \int_0^R {\rm d}r w(r) r^{2i+1}
\end{equation}
The ``diagonalized'' form of the partition function
\begin{equation} \label{3.21}
Z_N(\gamma) = \int {\cal D}\psi {\cal D}\xi
\prod_{i=0}^{\gamma(N-1)} \exp(\Xi_i w_i \Psi_i)
\end{equation}
implies that only correlators 
$\langle \Xi_{i_1} \Psi_{j_1} \Xi_{i_2} \Psi_{j_2} \cdots \rangle$ 
with $i_1+i_2+\ldots = j_1+j_2+\ldots$ will be nonzero.
The dependence of $Z_N(\gamma)$ on the set of moments
$\{ w_i \}_{i=0}^{\gamma(N-1)}$ is the crucial problem whose
solution would mean the complete exact solution 
(free energy, correlation functions) of the bulk 2D OCP
at the given coupling $\Gamma=2\gamma$.
Let us write down explicitly few examples of this dependence.
At $\gamma=1$, we have the simple result
\begin{equation} \label{3.22}
Z_N(1) = w_0 w_1 \cdots w_{N-1}
\end{equation}
At $\gamma=2$, using the anticommuting integral rules one finds
from (\ref{3.21}) for small particle numbers $N=2,3$ that
\begin{eqnarray}
Z_2(2) & = & w_0 w_2 + 2 w_1^2 \label{3.23} \\
Z_3(2) & = & w_0 w_2 w_4 + 2 w_0 w_3^2 + 2 w_1^2 w_4
+ 4 w_1 w_2 w_3 + 6 w_2^3 \label{3.24}
\end{eqnarray}
etc.
At $\gamma=3$ one has
\begin{eqnarray}
Z_2(3) & = & w_0 w_3 + 3^2 w_1 w_2 \label{3.25} \\
Z_3(3) & = & w_0 w_3 w_6 + 3^2 w_0 w_4 w_5 + 3^2 w_1 w_2 w_6
\nonumber \\ & & 
+ 6^2 w_1 w_3 w_5 + 15^2 w_2 w_3 w_4 \label{3.26}
\end{eqnarray}
etc.
There exists one model exactly solvable for every $\gamma$
and $N$, namely the 2D OCP constrained to a circle.
In that case $w(r) = \delta(r-1)/(2\pi)$ and, consequently,
$w_i=1$ for all $i=0,1,\ldots,\gamma(N-1)$.
It was proved in various ways \cite{Mehta} that
\begin{equation} \label{3.27}
Z_N(\gamma) = \frac{(\gamma N)!}{(\gamma!)^N N!}
\quad \quad \mbox{when all $w_i=1$}
\end{equation}  
Relations (\ref{3.22})-(\ref{3.26}) pass this test.

We now aim at analyzing the structure of a general summand
in $Z_N(\gamma)$.
It follows directly from the fermionic representation (\ref{3.21})
that each term is composed of just $N$ $w$'s,
$w_{i_1} w_{i_2} \cdots w_{i_N}$.
The transformation $\xi_i^{(\alpha)} \to \lambda^i \xi_i^{(\alpha)}$
for all $\alpha = 0,1,\ldots,\gamma$ indices and sites 
$i=0,1,\ldots,N-1$ implies $\Xi \to \lambda^i \Xi$.
As a consequence, the subscripts of the general term
$w_{i_1} w_{i_2} \cdots w_{i_N}$ must satisfy the relation
$\sum i = \gamma (0+1+\cdots+N-1) = \gamma N(N-1)/2$.
It is necessary to distinguish between $\gamma$ being an
odd or even integer.

For $\gamma$ an odd integer, the composite $\Xi$ and $\Psi$
operators are products of an odd number of anticommuting
variables and so they themselves satisfy the usual 
anticommutation rules
\begin{equation} \label{3.28}
\{ \Xi_i,\Xi_j \} = \{ \Psi_i,\Psi_j \} = \{ \Xi_i,\Psi_j \} = 0 
\end{equation} 
for all $i,j=0,1,\ldots,\gamma(N-1)$.
In particular it holds $\Xi_i^2 = \Psi_i^2 = 0$.
The expansion of each exponential in (\ref{3.21}) is thus
\begin{equation} \label{3.29}
\exp(\Xi_i w_i \Psi_i) = 1 + \Xi_i w_i \Psi_i
\end{equation}
and the partition function is represented as follows
\begin{eqnarray}
Z_N(\gamma) & = & \sum_{i_1<i_2<\ldots<i_N=0\atop
\left[\sum i = \gamma N(N-1)/2 \right]} C_{i_1,i_2,\ldots,i_N}^2
w_{i_1} w_{i_2} \cdots w_{i_N} \label{3.30} \\
C_{i_1,i_2,\ldots,i_N} & = & \int {\cal D}\xi~
\Xi_{i_1} \Xi_{i_2} \cdots \Xi_{i_N} \label{3.31}
\end{eqnarray}
We see that for odd $\gamma$ a given $w_i$ can occur in a summand
at most once.

When $\gamma$ is an even integer, the composite $\Xi$ and $\Psi$
operators are products of an even number of anticommuting variables
and so they commute with each other
\begin{equation} \label{3.32}
[\Xi_i,\Xi_j] = [\Psi_i,\Psi_j] = [\Xi_i,\Psi_j] = 0 
\end{equation} 
for all $i,j=0,1,\ldots,\gamma(N-1)$.
The expansion of an exponential is now a bit more complicated,
\begin{equation} \label{3.33}
{\rm e}^{\Xi_i w_i \Psi_i} = \left\{
\begin{array}{ll}
\sum_{j=0}^{[2i/\gamma]+1} \frac{1}{j!} (\Xi_i w_i \Psi_i)^j &
{\rm for}\ i=0,\ldots,\frac{\gamma}{2}(N-1) \cr
& \cr
\sum_{j=0}^{2N-1-[2i/\gamma]} \frac{1}{j!} (\Xi_i w_i \Psi_i)^j &
{\rm for}\ i=\frac{\gamma}{2}(N-1),\ldots,\gamma(N-1)
\end{array} \right.
\end{equation}
Next terms vanish because they contain second or higher power
of at least one anticommuting $\xi$ or $\psi$ variable.
Let us introduce the indices $\{ \alpha_i \}_{i=0}^{\gamma(N-1)}$
with the following value ranges:
\begin{equation} \label{3.34}
\alpha_i = \left\{
\begin{array}{ll}
0, 1,\ldots,\left[ \frac{2i}{\gamma} \right]+1 &
{\rm for}\ i=0,\ldots,\frac{\gamma}{2}(N-1) \cr
& \cr
0, 1,\ldots,2N-1-\left[ \frac{2i}{\gamma} \right] &
{\rm for}\ i=\frac{\gamma}{2}(N-1),\ldots,\gamma(N-1)
\end{array} \right.
\end{equation} 
The partition function for $\gamma$ an even integer is then
expressible as
\begin{equation} \label{3.35}
Z_N(\gamma) =  \sum_{\alpha_0,\ldots,\alpha_{\gamma(N-1)}
\atop \left[ \sum \alpha_i = N, \sum i\alpha_i =\frac{\gamma}{2}N(N-1)
\right]} C_{\alpha_0,\alpha_1,\ldots,\alpha_{\gamma(N-1)}}^2
\frac{w_0^{\alpha_0} w_1^{\alpha_1} \cdots 
w_{\gamma(N-1)}^{\alpha_{\gamma(N-1)}}}{\alpha_0! \alpha_1!
\cdots \alpha_{\gamma(N-1)}!} 
\end{equation}
where
\begin{equation} \label{3.36}
C_{\alpha_0,\alpha_1,\ldots,\alpha_{\gamma(N-1)}}
= \int {\cal D}\xi~ \Xi_0^{\alpha_0} \Xi_1^{\alpha_1} \cdots
\Xi_{\gamma(N-1)}^{\alpha_{\gamma(N-1)}} 
\end{equation}

The underlying fermionic representations, relations (\ref{3.30}), 
(\ref{3.31}) for $\gamma$ odd and relations (\ref{3.34})-(\ref{3.36})
for $\gamma$ even, contain the unknown coefficients which can be
formally written in both cases as
\begin{equation} \label{3.37}
C_{i_1,\ldots,i_N} = \int {\cal D}\xi~
\Xi_{i_1} \Xi_{i_2} \cdots \Xi_{i_N}, \quad \quad
\sum i = \frac{\gamma}{2} N (N-1)
\end{equation}
The ``basic sector'' of indices is
\begin{equation} \label{3.38}
\begin{array}{ll}
i_1<i_2<\ldots<i_N, &  \mbox{for $\gamma$ odd} \cr
i_1\le i_2\le \ldots\le i_N, &  \mbox{for $\gamma$ even}
\end{array}
\end{equation}
The $C$-coefficient with an arbitrary sequence of indices can be expressed
in terms of the basic one with indices ordered according to (\ref{3.38})
through a successive exchange of nearest-neighbor couples of composite
operators,
\begin{equation} \label{3.39}
C_{i_1,\ldots,i_j,i_{j+1},\ldots,i_N} = (-1)^{\gamma}
C_{i_1,\ldots,i_{j+1},i_j,\ldots,i_N} 
\end{equation}
It is evident that
\begin{equation} \label{3.40}
C_{0,\gamma,\ldots,\gamma(N-1)} = 1
\end{equation} 

One can evaluate the $C$-coefficients directly from their 
definition (\ref{3.37}) by using the rules of integration over
anticommuting variables.
However, there exists another simpler way how to determine
these coefficients.
Let $P_{\gamma-1}(k,k')$ be a general polynomial of the $(\gamma-1)$
order in the $k,k'$ variables, and consider this polynomial in the 
following combination with the composite operators
\begin{equation} \label{3.41}
\sum_{k,k'=0\atop (k+k'=K)}^{\gamma(N-1)}
P_{\gamma-1}(k,k') \Xi_k \Xi_{k'}; \quad \quad
K=0,1,\ldots,2\gamma(N-1)
\end{equation}
Representing the $\Xi$-operators in terms of the anticommuting
$\xi$-variables [see relation (\ref{3.13})], Eq. (\ref{3.41})
can be rewritten as
\begin{equation} \label{3.42}
\sum_{{k_1,\ldots,k_{\gamma}\atop k'_1,\ldots,k'_{\gamma}}=0}^{N-1}
P_{\gamma-1}(k_1+\cdots+k_{\gamma},k'_1+\cdots+k'_{\gamma})
\xi_{k_1}^{(1)}\cdots \xi_{k_{\gamma}}^{(\gamma)}
\xi_{k'_1}^{(1)}\cdots \xi_{k'_{\gamma}}^{(\gamma)}
\delta_{K,k_1+\cdots k_{\gamma}+k'_1+\cdots k'_{\gamma}}
\end{equation}
Since $P_{\gamma-1}$ is the polynomial of the $(\gamma-1)$ order
in its arguments, at least one couple of the $k,k'$ factors
with the same subscript, say the couple $k_1$ and $k'_1$ 
associated with $\xi^{(1)}$, does not occur in the given term
of the expansion of 
$P_{\gamma-1}(k_1+\cdots k_{\gamma},k'_1+\cdots k'_{\gamma})$.
Thus, since
\begin{eqnarray}
& & \sum_{k_1,k'_1=0}^{N-1} \xi_{k_1}^{(1)} \xi_{k'_1}^{(1)}
\delta_{K,k_1+\cdots k_{\gamma}+k'_1+\cdots k'_{\gamma}} \nonumber \\
& &  = \sum_{k_1<k'_1} 
\left( \xi_{k_1}^{(1)} \xi_{k'_1}^{(1)}+\xi_{k'_1}^{(1)} \xi_{k_1}^{(1)}
\right) \delta_{K,k_1+\cdots k_{\gamma}+k'_1+\cdots k'_{\gamma}} = 0
\label{3.43}
\end{eqnarray}
the polynomial combination with composite operators (\ref{3.41})
vanish.
Passing from $P_{\gamma-1}(k,k')$ to $Q_{\gamma-1}(k+k',k-k')$
one has
\begin{equation} \label{3.44}
\sum_{k,k'=0\atop (k+k'=K)}^{\gamma(N-1)}
(k-k')^n \Xi_k \Xi_{k'} = 0
\quad \quad \mbox{for $n=0,1,\ldots,\gamma-1$}
\end{equation}
We remind that the $\Xi$-operators anticommute for $\gamma$ odd
and commute for $\gamma$ even. 
This is why for $\gamma$ odd only odd powers of $(k-k')$ provide 
a nontrivial information and for $\gamma$ even only even powers 
of $(k-k')$ are informative.
In view of the representation (\ref{3.37}), we finally arrive at 
a homogeneous set of linear equations for the coefficients $C$:

\noindent if $\gamma$ is odd,
\begin{equation} \label{3.45}
\sum_{k,k'=0\atop (k+k'=K)}^{\gamma(N-1)}
(k-k')^n C_{i_1,\ldots,i_{N-2},k,k'} = 0; \quad \quad
n= 1,3,\ldots,\gamma-2
\end{equation}
if $\gamma$ is even,
\begin{equation} \label{3.46}
\sum_{k,k'=0\atop (k+k'=K)}^{\gamma(N-1)}
(k-k')^n C_{i_1,\ldots,i_{N-2},k,k'} = 0; \quad \quad
n= 0,2,\ldots,\gamma-2
\end{equation}
In both cases $K$ runs over $0, 1,\ldots, 2\gamma(N-1)$.

In the first nontrivial case of $\gamma=2$, according to Eq. (\ref{3.39}) 
the interchange of subscripts does not change the sign of the 
$C$-coefficients.
The normalization is $C_{0,2,\ldots,2(N-1)}=1$.
Formula (\ref{3.46}) with $n=0$ applies.
For $N=2$, the normalization $C_{0,2}=1$ is supplemented by the 
only equation
\begin{equation} \label{3.47}
0 = C_{0,2} + C_{1,1} + C_{2,0}
\end{equation}
so that $C_{1,1}=-2$. 
Inserting the two coefficients into the representation (\ref{3.35})
one recovers the exact relation (\ref{3.23}).
For $N=3$, the normalization $C_{0,2,4}=1$ is supplemented by the
(overcomplete) set of five equations
\begin{eqnarray}
0 & = & C_{0,2,4}+C_{0,3,3}+C_{0,4,2} \nonumber \\
0 & = & C_{1,1,4}+C_{1,2,3}+C_{1,3,2}+C_{1,4,1} \nonumber \\
0 & = & C_{2,0,4}+C_{2,1,3}+C_{2,2,2}+C_{2,3,1}+C_{2,4,0} \label{3.48}\\
0 & = & C_{3,0,3}+C_{3,1,2}+C_{3,2,1}+C_{3,3,0} \nonumber \\
0 & = & C_{4,0,2}+C_{4,1,1}+C_{4,2,0} \nonumber
\end{eqnarray}
which implies
$C_{0,3,3}=C_{1,1,4}=-2$, $C_{1,2,3}=2$ and $C_{2,2,2}=-6$.
Inserting these coefficients into (\ref{3.35}) one recovers
the relation (\ref{3.24}).
We suggest that the sets of linear equations, (\ref{3.45}) for
$\gamma$ odd and (\ref{3.46}) for $\gamma$ even,  constitute, 
together with the relation (\ref{3.39}) and the normalization (\ref{3.40}), 
complete (more precisely, overcomplete) sets to be solved for 
the $C$-coefficients.
We have checked this conjecture for $\gamma=2$ $(\Gamma=4)$ up to $N=13$
particles and for $\gamma=3$ $(\Gamma=6)$ up to $N=9$ particles.

In comparison with the technique of anticommuting variables, 
the algorithms for solving a set of linear equations are much
faster.
The presented scheme is therefore very convenient for exact
computer calculations of the plasma with a finite number of particles.
From the methods dealing with the finite number $N$ of particles
\cite{Dunne}-\cite{Tellez}, the closest one \cite{Tellez} also
provides the representations of the partition function (\ref{3.30})
for $\gamma$ odd and (\ref{3.35}) for $\gamma$ even.
The $C$-coefficients are expressed there as multiple integrals
over the unit circle, and consequently as multiple summations
over all possible permutations of $N$ indices.
Such algorithm is inferior to the present approach.

At this stage, we were not able to find a simplification of 
the underlying sets of linear equations in the thermodynamic limit.
In spite of this the fact that the crucial $C$-coefficients
are determined by linear equations may indicate
the exact solvability of the 2D OCP at $\Gamma=2*$integer.

\renewcommand{\theequation}{4.\arabic{equation}}
\setcounter{equation}{0}

\section{GAUGE INVARIANCE}
In the case of classical statistical systems with short-ranged interactions
among constituents, the thermodynamic limit of an intensive quantity does 
not depend in general on the shape of the system and on the conditions at 
the boundary given by the surrounding medium.
This may be no longer true for macroscopic systems with
long-ranged interactions.
A typical example is the shape-dependence of the dielectric susceptibility 
tensor for Coulomb conductors \cite{Landau} caused by the long-range
decay of charge correlations along the boundary
\cite{Choquard2,Jancovici7}.
Gauge invariance of Coulomb fluids is another, in a certain sense opposite, 
phenomenon related to the long-range nature of particle interactions.
Let us consider a macroscopic Coulomb conductor in an arbitrary
shaped domain $\Lambda$, with perhaps some charge distribution on 
the boundary $\partial \Lambda$.
The effect of the boundary is then reflected in the bulk through 
a long-ranged one-body potential whose Laplacian is zero inside 
the domain $\Lambda$.  
The assumption of gauge invariance states that this one-body ``gauge''
potential is perfectly screened by the Coulomb system at macroscopic
distances from the boundary, and so it does not affect 
the averaged particle distributions in the bulk interior.

The above developed fermionic formalism will be now used
to prove gauge invariance of macroscopic Coulomb fluids
at the special value of the coupling $\Gamma=2$.
The original proof of gauge invariance (up to the gauge potential
being the polynomial of degree 2 in spatial $x$ and $y$ components) 
at $\Gamma=2$ was presented in Ref. \cite{Cornu} for
a class of possibly inhomogeneous backgrounds.
The present proof is an alternative one, applicable also to
non-physical situations with a complex gauge potential. 

At $\Gamma=2$ and in the units of $\pi n_b=1$, the one-body
Boltzmann factor of the circularly symmetric background (\ref{3.9})
defined in an infinite space reads
\begin{equation} \label{4.1}
w(z,{\bar z}) = \frac{1}{\pi} {\rm e}^{-z{\bar z}},
\quad \quad \Lambda = R^2
\end{equation}
The $N\times N$ matrix (\ref{3.14}) becomes diagonal,
$w_{ij} = w_i \delta_{ij}$, with elements
\begin{equation} \label{4.2}
w_i = 2 \int_0^{\infty} {\rm d}r {\rm e}^{-r^2} r^{2i+1}
= i!; \quad \quad i = 0,1,\ldots,N-1
\end{equation}  
Combining Eqs. (\ref{3.15}) and (\ref{3.18}), the latter written
as $\langle \xi_i \psi_j \rangle = \delta_{ij}/w_i$, the particle
density at distance $r$ from the center is given by
\begin{equation} \label{4.3}
\frac{n(r)}{n_b} = {\rm e}^{-r^2} \sum_{i=1}^{N-1}
\frac{r^{2i}}{i!}
\end{equation}
In the limit $N\to\infty$ and for finite $r$, the bulk particle
density is constant, $n(r)=n_b$, as was expected.

Let us now consider the general case of a homogeneous background
characterized by the one-body Boltzmann factor (\ref{3.10}) with
gauge degrees of freedom,
\begin{equation} \label{4.4}
w(z,{\bar z}) = \frac{1}{\pi}
\exp\left[ - z{\bar z} - \sum_{s=1}^{\infty}
( a_s z^s + b_s {\bar z}^s ) \right]
\end{equation}
The interaction (\ref{3.14}) is non-diagonal,
\begin{equation} \label{4.5}
w_{ij} = \int \frac{{\rm d}^2 z}{\pi} z^i {\bar z}^j
\exp\left[ - z{\bar z} - \sum_s ( a_s z^s + b_s {\bar z}^s ) \right]
\end{equation}
$i,j=0,1,\ldots,N-1$.
The strength of gauge parameters $\{ a_s,b_s \}$
must be such that integrals in (\ref{4.5}) converge.
For a real gauge potential with $a_s={\bar b}_s=a{\rm e}^{{\rm i}\phi}$, 
writing $z=r{\rm e}^{{\rm i}\varphi}$ we have
$a_s z^s + b_s {\bar z}^s = 2 a r^s \cos(\phi + s\varphi)$.
If $s\ge 3$, for a fixed $\phi$ there always exist the $\varphi$-angles
such that $a \cos(\phi + s\varphi)<0$, which implies divergent
$r\to\infty$ contributions to the integral in (\ref{4.5}).
Thence all gauge coefficients $\{ a_s,b_s \}$ with $s\ge 3$ 
must be equal to zero in physically acceptable situations.
For $s=2$, the convergence is ensured provided that $2 a < 1$.   

In view of Eqs. (\ref{3.15}) and (\ref{3.18}), the particle density
at point $(z,{\bar z})$ is given by
\begin{equation} \label{4.6}
\frac{n(z,{\bar z})}{n_b} = \pi w(z,{\bar z})
\sum_{i,j=0}^{N-1} w_{ji}^{-1} z^i {\bar z}^j
\end{equation}
In the thermodynamic limit $N\to\infty$, the bulk particle density
is expected to be constant, $n(z,{\bar z}) = n_b$.
This is true iff
\begin{equation} \label{4.7}
\sum_{i,j=0}^{\infty} w_{ji}^{-1} z^i {\bar z}^j
= \exp\left[ z{\bar z} + \sum_s ( a_s z^s + b_s {\bar z}^s ) \right]
\end{equation}
This equation can be understood as the generating relation, 
with $z$ and ${\bar z}$ assumed as independent variables, 
for the inverse elements of the infinite $w$-matrix with elements (\ref{4.5}).

To prove that the quantities $w_{ji}^{-1}$ determined by Eq. (\ref{4.7})
form indeed the inverse matrix of the one with elements (\ref{4.5})
we have to show that it holds
\begin{equation} \label{4.8}
\sum_{k=0}^{\infty} w_{ik} w_{kj}^{-1} = \delta_{ij} , \quad \quad
\sum_{k=0}^{\infty} w_{ik}^{-1} w_{kj} = \delta_{ij}
\end{equation}
for all $i,j=0,1,\ldots$.
Let us consider the auxiliary function
\begin{equation} \label{4.9}
K_i(t) = \sum_{j,k=0}^{\infty} 
w_{ik} w_{kj}^{-1} t^j
\end{equation}
Presuming the validity of the generating Eq. (\ref{4.7}),
a series of straightforward transformations yields
\begin{eqnarray}
K_i(t) & = & \sum_{j,k=0}^{\infty} \int \frac{{\rm d}^2 z}{\pi}
z^i {\bar z}^k w(z,{\bar z}) w_{kj}^{-1} t^j \nonumber \\
& = & \int \frac{{\rm d}^2 z}{\pi} z^i
\exp\left[ -z{\bar z} + t{\bar z} - \sum_s a_s (z^s-t^s) \right]
\label{4.10} \\
& = & \int \frac{{\rm d}^2 z}{\pi} (z+t)^i
\exp\left\{ -z{\bar z} - {\bar t} z - \sum_s a_s 
\left[(z+t)^s-t^s \right] \right\} \nonumber
\end{eqnarray}
Since $(z+t)^s-t^s=z^s+s t z^{s-1}+\cdots +st^{s-1}z$, 
only terms $t^i$ and $-z{\bar z}$ survive
from $(z+t)^i$ and the exponential, respectively.
Consequently,
\begin{equation} \label{4.11}
K_i(t) = t^i \quad \quad \mbox{for all $i=0,1,\ldots$}
\end{equation}
With regard to the definition (\ref{4.9}) of $K_i(t)$,
this equation implies the first relation in Eq. (\ref{4.8}).
The second relation can be obtained in a similar way
by showing that
$L_i({\bar t}) = \sum_{j,k=0}w_{ik}^{-1} w_{kj} {\bar t}^j$
is equal to ${\bar t}^i$ for all $i=0,1,\ldots$.
The proof is accomplished.

One can readily show that the proof of gauge invariance for the
one-body density automatically ensures gauge
invariance for the two-body density (\ref{3.16}) with
correlators (\ref{3.19}).

Strictly speaking, gauge invariance has meaning only for
real gauge potentials; the short discussion after formula
(\ref{4.5}) tells us that only gauge potential being polynomial
of degree 2 in $z$ and ${\bar z}$ complex coordinates is
allowed in physical situations.
On the other hand, the proof of the matrix inversion
(\ref{4.8})-(\ref{4.11}) requires only the finite values
of matrix elements (\ref{4.5}), without putting any further
restriction on the $\{ a_s,b_s \}$ gauge parameters which
may therefore be unphysical.
Typical unphysical examples leading to finite values of matrix
elements are $b_s=0$ or $a_s=b_s={\rm i}$. 
It is interesting from a mathematical point of view that
there exists a large family of infinite matrices, 
with elements $w_{ij}$ $(i,j=0,1,\ldots)$ defined by Eq. (\ref{4.5}), 
which are explicitly invertible by using
the closed-form generating formula (\ref{4.7}).
A detailed analysis of this mathematical peculiarity
goes beyond the scope of the present paper.

\section{Conclusion}
In the canonical ensemble, the 2D jellium is equivalent to
the 2D Euclidean-field theory with the action (\ref{2.33}).
Here, the divergent Coulomb self-energy does not renormalize
the model's parameters like it is in the sine-Gordon representation
of the 2D Coulomb gas.
In contrast to the sine-Gordon model, the quantum analogue
of the present Euclidean theory is conjectured
to be not integrable on the classical
level (only such realizations of the $\phi$-field are considered
which minimalize the action) due to the lack of an infinite 
sequence of integrals of motion.
The classical non-integrability does not exclude 
the complete quantum (all realizations of the $\phi$-field are considered) 
integrability at specific discrete values of $\Gamma$, 
like it is at the free-fermion 
$\Gamma=2$ point.
This free-fermion coupling belongs to a family of couplings 
$\Gamma=2\gamma$ ($\gamma$ integer) which admit a 1D fermionic 
representation of the partition function.
The fermionic representations, relations (\ref{3.30}) and
(\ref{3.31}) for $\gamma$ odd and relations 
(\ref{3.34})-(\ref{3.36}) for $\gamma$ even, contain the
unknown $C$-coefficients.
These coefficients are determined by the homogeneous sets of linear
equations, (\ref{3.45}) for $\gamma$ odd and (\ref{3.46}) for
$\gamma$ even, supplemented by the exchange formula (\ref{3.39})
and the normalization (\ref{3.40}).
This feature is a sign of integrability.
The present analysis might be a challenge for specialists in 
the Field Theory.

The proof of gauge invariance of the 2D OCP at $\Gamma=2$
is related to the exact inversion of a class of infinite-dimensional
matrices which elements are determined by non-Gaussian integrals (\ref{4.5}).
This is interesting from a mathematical point of view. 

\section*{Acknowledgements}
Section 2 aroused from stimulating discussions with B. Jancovici
whom I am also thank for careful reading of the manuscript and
very useful comments.
I am grateful to P. Kalinay for computer calculations in the symbolic
language {\it Reduce}.
The support by a VEGA grant is acknowledged.


\newpage

\begin{thebibliography}{9}

\bibitem{Martin} Ph. A. Martin,
{\it Rev. Mod. Phys.} {\bf 60}:1075 (1988).

\bibitem{Blum} L. Blum, C. Gruber, J. L. Lebowitz, and Ph. A. Martin,
{\it Phys. Rev. Lett.} {\bf 48}:1769 (1982).

\bibitem{Jancovici1} B. Jancovici,
{\it J. Phys.: Condens. Matter} {\bf 14}:9121 (2002),
and references quoted there.

\bibitem{Torres} A. Torres and G. T\'ellez,
{\it J. Phys. A} {\bf 37}:2121 (2004).

\bibitem{Jancovici2} B. Jancovici,
in {\it Inhomogeneous Fluids}, D. Henderson, ed. 
(Dekker, New York, 1992), pp. 201-237.

\bibitem{Forrester1} P. J. Forrester, 
{\it Phys. Rep.} {\bf 301}:235 (1998).

\bibitem{Salzberg} A. Salzberg and S. Prager,
{\it J. Chem. Phys.} {\bf 38}:2587 (1963).

\bibitem{Samaj1} L. {\v S}amaj,
{\it J. Phys. A} {\bf 36}:5913 (2003).

\bibitem{Ginibre} J. Ginibre,
{\it J. Math. Phys.} {\bf 6}:440 (1965).

\bibitem{Prange} R. E. Prange and S. M. Girvin,
{\it The Quantum Hall Effect} (Springer, New York, 1987).

\bibitem{Francesco} P. Di Francesco, M. Gaudin, C. Itzykson,
and F. Lesage,
{\it Int. J. Mod. Phys. A} {\bf 9}:4257 (1994).

\bibitem{Choquard1} Ph. Choquard and J. Cl\'erouin,
{\it Phys. Rev. Lett.} {\bf 50}:2086 (1983).

\bibitem{Moore} M. A. Moore and A. P\'erez-Garrido,
{\it Phys. Rev. Lett.} {\bf 82}:4078 (1999).

\bibitem{Jancovici3} B. Jancovici,
{\it Phys. Rev. Lett.} {\bf 46}:386 (1981).

\bibitem{Vieillefosse} P. Vieillefosse and J. P. Hansen,
{\it Phys. Rev. A} {\bf 12}:1106 (1975).

\bibitem{Kalinay1} P. Kalinay, P. Marko{\v s}, L. {\v S}amaj,
and I. Trav{\v e}nec,
{\it J. Stat. Phys.} {\bf 98}:639 (2000).

\bibitem{Forrester2} P. J. Forrester,
{\it J. Stat. Phys.} {\bf 63}:491 (1991).

\bibitem{Jancovici4} B. Jancovici, G. Manificat, and C. Pisani,
{\it J. Stat. Phys.} {\bf 76}:307 (1994).

\bibitem{Jancovici5} B. Jancovici and E. Trizac,
{\it Physica A} {\bf 284}:241 (2000).

\bibitem{Dunne} G. V. Dunne,
{\it Int. J. Mod. Phys. B} {\bf 7}:4783 (1994).

\bibitem{Scharf} T. Scharf, J. Y. Thibon, and B. G. Wybourne,
{\it J. Phys. A} {\bf 27}:4211 (1994).

\bibitem{Samaj2} L. {\v S}amaj, J. K. Percus, and M. Koles\'{\i}k,
{\it Phys. Rev. E} {\bf 49}:5623 (1994).

\bibitem{Tellez} G. T\'ellez and P. J. Forrester,
{\it J. Stat. Phys.} {\bf 97}:489 (1999).

\bibitem{Samaj3} L. {\v S}amaj and J. K. Percus,
{\it J. Stat. Phys.} {\bf 80}:811 (1995).

\bibitem{Samaj4} L. {\v S}amaj, P. Kalinay, and I. Trav{\v e}nec,
{\it J. Phys. A} {\bf 31}:4149 (1998).

\bibitem{Ghoshal} S. Ghoshal and A. B. Zamolodchikov,
{\it Int. J. Mod. Phys. A} {\bf 9}:3841 (1994).

\bibitem{Cornu} F. Cornu, B. Jancovici, and L. Blum,
{\it J. Stat. Phys.} {\bf 50}:1221 (1988).

\bibitem{Gradshteyn} I. S. Gradshteyn and I. M. Ryzhik,
{\it Tables of Integrals, Series, and Products},
5th edn. (Academic Press, London, 1994). 

\bibitem{Brydges} D. C. Brydges and Ph. A. Martin,
{\it J. Stat. Phys.} {\bf 96}:1163 (1999).

\bibitem{Lieb} E. H. Lieb and H. Narnhofer,
{\it J. Stat. Phys.} {\bf 12}:291 (1975);
ibid {\bf 14}:465 (1976).

\bibitem{Fantoni} R. Fantoni, B. Jancovici, and G. T\'ellez,
{\it J. Stat. Phys.} {\bf 112}:27 (2003).

\bibitem{Minnhagen} P. Minnhagen,
{\it Rev. Mod. Phys.} {\bf 59}:1001 (1987).

\bibitem{Brilliantov} N. V. Brilliantov,
{\it Contrib. Plasma Phys.} {\bf 38}:489 (1998).

\bibitem{Zinn} J. Zinn-Justin,
{\it Quantum Field Theory and Critical Phenomena}
(Clarendon Press, Oxford, 1999), 3rd edn.

\bibitem{Dodd} R. K. Dodd and R. K. Bullough,
{\it Proc. R. Soc. London A} {\bf 352}:481 (1977).

\bibitem{Samaj5} L. {\v S}amaj,
{\it J. Stat. Phys.} {\bf 111}:261 (2003).

\bibitem{Forrester3} P. J. Forrester and B. Jancovici,
{\it Int. J. Mod. Phys. A} {\bf 5}:941 (1996).

\bibitem{Jancovici6} B. Jancovici,
{\it J. Stat. Phys.} {\bf 28}:43 (1982).

\bibitem{Berezin} F. A. Berezin,
{\it The method of Second Quantization}
(Academic Press, New York, 1966).

\bibitem{Mehta} M. L. Mehta,
{\it Random Matrices}, 2nd edn.
(Academic, London, 1990).

\bibitem{Landau} L. D. Landau and E. M. Lifshitz,
{\it Electrodynamics of Continuous Media}, 2nd edn.
(Pergamon, Oxford, 1963), Chapter I.

\bibitem{Choquard2} Ph. Choquard, B. Piller, R. Rentsch, and P. Vieillefosse,
{\it J. Stat. Phys.} {\bf 55}:1185 (1989).

\bibitem{Jancovici7} B. Jancovici and L. {\v S}amaj,
{\it J. Stat. Phys.} {\bf 114}:1211 (2004).

\end{thebibliography}
\end{document}